\documentclass[%
 reprint,
 amsmath,amssymb,
 aps,
 prb,  % PRL Style
 superscriptaddress, % Change the address style to superscripts
 longbibliography,
]{revtex4-2}

\usepackage{graphicx}% Include figure files
\usepackage{dcolumn}% Align table columns on decimal point
\usepackage{bm}% bold math
\emergencystretch=\maxdimen
\begin{document}
\preprint{APS/123-QED}

\title{Exploration of Wire Array Metamaterials for the Plasma Axion Haloscope}
\author{M. Wooten}
\affiliation{Department of Nuclear Engineering, University of California Berkeley,
 Berkeley, CA 94720 USA}
\author{A. Droster}
\affiliation{Department of Nuclear Engineering, University of California Berkeley,
 Berkeley, CA 94720 USA}
\author{Al Kenany}
\affiliation{Department of Nuclear Engineering, University of California Berkeley,
 Berkeley, CA 94720 USA}
\affiliation{Accelerator Technology and Applied Physics Division, Lawrence Berkeley National Laboratory, Berkeley, CA 94720}
\author{D. Sun}
\affiliation{Department of Nuclear Engineering, University of California Berkeley,
 Berkeley, CA 94720 USA}
\author{S.M. Lewis}
\affiliation{Department of Nuclear Engineering, University of California Berkeley,
Berkeley, CA 94720 USA}
\author{K. van Bibber}
\affiliation{Department of Nuclear Engineering, University of California Berkeley,
 Berkeley, CA 94720 USA}

\date{\today}% It is always \today, today,
             %  but any date may be explicitly specified

\begin{abstract}
A plasma haloscope has recently been proposed as a feasible approach to extend the search for dark matter axions above 10 GHz ($\sim$ 40 \textmu eV), whereby the microwave cavity in a conventional axion haloscope is supplanted by a wire array metamaterial. Since the plasma frequency of a metamaterial is determined by its unit cell, and is thus a bulk property, a metamaterial resonator of any frequency can be made arbitrarily large, in contrast to a microwave cavity which incurs a steep penalty in volume with increasing frequency. To assess the actual potential of this concept as a practical dark matter haloscope, the basic properties of wire array metamaterials have been investigated through an extensive series of $S_{21}$ measurements in the 10 GHz range.  This report presents some new systematics of wire array metamaterials themselves including the approach to full plasmonic behavior, the applicability of the semianalytic theory of Belov, and estimates of the loss term which bode favorably for the plasmonic haloscope application. This present work constitutes the first precision test of the semianalytic theory of Belov et al., for which the predicted plasma frequency agrees with the experimental value at the 0.1\% level.

\end{abstract}

%\keywords{Suggested keywords}%Use showkeys class option if keyword
                              %display desired
\maketitle

%\tableofcontents

\section{\label{sec:level1}Introduction}
The axion, a light pseudoscalar emerging from the Peccei-Quinn mechanism to suppress CP-violating effects in the Strong Interaction [1-3] would also be copiously produced in the early universe, and thus represents an interesting dark matter candidate. The mass of the axion corresponding to the estimated dark matter density of the universe, however is largely unconstrained over eleven orders of magnitude. In cosmological scenarios of long low-scale inflation, the axion mass may be as low as 10\textsuperscript{-12} eV [4,5]. In the case of Peccei-Quinn symmetry breaking after inflation, the mass is much higher, but more significantly is uniquely determined and in principle exactly calculable [6,7]. Calculating the evolution of the axion string network over cosmological time is an extremely challenging problem. However, recent adaptive mesh refinement simulations represent a major advancement over prior work and now predict the axion mass to be in the range (40-180) {\textmu}eV with a preferred value of 65 ± 6 {\textmu}eV [7]; larger simulations will constrain this range more tightly in the near future.

Along with the many orders of magnitude of mass to be searched, the couplings of the axion to radiation and matter are also expected to be extraordinarily weak, making the experimental front extremely challenging as well. The primary strategy for the detection of halo dark matter axions is their conversion to photons in a magnetic field, resonantly enhanced inside a microwave cavity [8,9]. Other concepts are also under development [10,11]. The predicted conversion power is miniscule for current experiments, $ \sim 10^{–(23-24)}$ Watts [12-15], and thus rendering such a signal observable depends on two factors. The first is aggressive noise reduction, where the microwave cavity axion search has been both a driver and beneficiary of advances in quantum metrology [15,16]. The second is maximizing the signal power itself, where the conversion power scales with factors under the experimenter's control as $P \propto B_{0}^{2}VCQ$ , where $B_{0}$ is the strength of the magnetic field. The remaining three parameters are properties of the microwave cavity, i.e. the volume \emph{V}, quality factor \emph{Q}, and form factor \emph{C} for the operating mode, the latter reflecting the overlap of the mode’s \emph{E}-field with the external magnetic field $B_{0}$. The $\rm{TM_{010}}$ is generally the mode of choice, as its form factor is usually the largest. The connection between frequency and physical size of a microwave cavity as limited by the magnet bore, is what in fact centers current experiments around the 0.5-5 GHz decade, or ~ 2-20 \textmu eV  (1 GHz = 4.136 \textmu eV). To reach axions of much lower mass, neV or lower, associated with slow-roll inflation scenarios, variants of the microwave cavity scheme have been proposed whereby the frequency of the resonator is determined by a lumped-element LC circuit external to the conversion volume [17,18]. A pilot experiment based on this concept has been carried out [19], and a full-scale experiment capable of reaching QCD axion models is being actively pursued [20].

For the post-inflation axion, the search will need resonators that can scan up to an order of magnitude higher in frequency than current experiments. A single miniature cavity has been used to scan in the mass region of the post-inflation axion; owing to its small volume however, it was not sensitive to QCD axions, but set limits on axion-like particles, or ALPS [21]. The practicality of scaling current cylindrical cavities to this range is limited due to the inherent connection between a cavity's frequency and its size, which would substantially reduce signal power. Various strategies have been pursued to make a resonator that is simultaneously large volume and high frequency. One is to insert multiple axial metal posts into a single large cavity to raise and tune its frequency [22]. Another is to make an array of multiple small cavities all tuned in concert and combined in phase [23]. Yet another utilizes a series of partitions within the cavity, subdividing it into strongly coupled subvolumes [24,25,26].  An entirely different approach is an open resonator with dielectric slabs of variable spacing [27]. But even these are very complex from an engineering standpoint, and are limited to only modest extensions of currently accessible frequency for the microwave cavity axion search.

\subsection{The Tunable Plasma Haloscope}

Motivated by the need to develop a practical search strategy for the post-inflation axion, Lawson et al. have proposed a radically different approach to decoupling the size and frequency of a resonator by replacing the microwave cavity with a metamaterial serving a surrogate plasma [28]. Whereas the frequency of a cavity mode is determined by its boundary conditions and thus its physical dimensions, for a metamaterial, the plasma frequency is a function of its unit cell, and is thus independent of its physical dimensions. In principle a resonator could be designed to be for all practical purposes both arbitrarily high in frequency and arbitrarily large. This would open up new design options, for example an experiment based on a very large volume, medium-field magnet such as a whole-body MRI scanner, or a collider detector magnet, in contrast to current experiments built on small volume, very high field magnets. Lawson et al. propose a metamaterial of a uniform array of thin parallel wires as a potentially viable solution, for which there is a well-developed theory and some experimental validation [29-31].   

The topic of axion-photon mixing in plasmas has a long history going back to the original work of Raffelt and Stodolsky [32]. The utility of the resonant conversion condition when the plasma frequency equals the axion mass has been explored in several contexts, including producing beams of axion-like particles with laser wakefield accelerators [33], and extending the mass reach of axion helioscopes [34], the latter application having in fact been reduced to practice in all magnetic helioscopes, e.g. [35]. Plasma mediated conversion of dark matter axions in the magnetospheres of neutron stars has also been studied theoretically [36,37], and one such radio telescope search has been carried out [38]. The purpose of our investigation is to provide a thorough experimental assessment of the real feasibility of wire array metamaterials for the axion haloscope application. This first report focuses on measuring the basic properties of wire array metaterials, and compare with calculations. Detailed measurements have elucidated some features of these metamaterials not previously recognized, particularly the approach to asymptotic behavior, the excellent agreement with the semianalytical theory, and provide an estimate of the resonator quality factor, \emph{Q} when the diameter and spacing of the wires are optimized at a given frequency. The second report will focus primarily on the dependence of the plasma frequency as a function of the properties of the unit cell of the array, i.e. its geometrical configuration, which has a more practical aim of demonstrating a usable dynamic range in frequency in an actual search.

As a simple illustrative case, Pendry et al. [29] consider an array of thin wires of radius \emph{r} in a square lattice of separation \emph{a}, oriented in the $\hat{z}$ direction, and in an approximate treatment demonstrate that the array supports a longitudinal plasmonic mode, with a plasma frequency given by

\begin{equation}
\nu_{p}^{2} = \frac{ c^{2}}{2\pi a^{2} \rm{ln}(\frac{a}{r})}.
\end{equation}

For \emph{a} = 5 mm and \emph{r} = 25 {\textmu}m, representative of parameters in our study, the plasma frequency, $\nu_{p}$ = 10.75 GHz.  Belov et al. have developed a more accurate analytic result and have also generalized it to the case of a rectangular lattice of dimensions a, b [30].

\begin{subequations}
\label{eq:whole}
\begin{equation}
\nu_{p}^{2} = \frac{c^{2}}{2\pi s^{2}\left( \rm{ln}\left(\frac{s}{2\pi r}\right) +F(\mathit{u})\right)},
\end{equation}
\rm{where} $\mathit{s} = \sqrt{\mathit{ab}}, \mathit{u} = \mathit{a}/\mathit{b}, \rm{and}$
\begin{equation}
\rm{F}(u) = -\frac{1}{2}\rm{ln}(\mathit{u}) + \sum_{n=1}^{\infty} \frac{\rm{coth}(\pi n\mathit{u}) - 1}{n} + \frac{\pi \mathit{u}}{6} 
\end{equation}
\end{subequations}

\section{\label{sec:level2}Methods}
\subsection{Goals and Approach}
The substitution of a wire array metamaterial for a microwave cavity is not entirely straightforward; there are features that differ from a cavity resonator that affect the practical considerations in building a metamaterial-based haloscope. These include, for example, the design of the antenna to couple out power to the receiver. Furthermore, while the plasmon excitation is largely self-contained, the wire array metamaterial itself needs to be shielded by a conducting enclosure, that that has to be incorporated into the microwave simulations of the overall resonator. This paper will be limited just to basic studies of the wire array metamaterial that will determine its fundamental suitability for the haloscope application. These include the dependence of its properties as a function of its size, some estimate of \emph{Q} that may be achievable in situ, and the applicability of the semianalytic theory as a tool to guide the design of an actual haloscope resonator.

Measurement of the microwave transmission, i.e. the $S_{21}$ parameter is an accurate and reliable tool for determining the properties of a complex dielectric, which has been previously applied in early studies of wire array metamaterials [29,31]. The $S_{21}$ is one of the four scattering parameters for a two-port device, specifically the voltage ratio of the output of the second port to the input of the first port, i.e.  $S_{21} = {V_2}^{+} / {V_1}^{+}$. Thus $S_{21}$ is complex number, of magnitude less than or equal to 1 for a device without gain. Furthermore, particularly to study the evolution of metamaterial behavior as a function of spatial extent, it was natural and convenient to assemble the array as a series of wire frames which could be stacked with various spacings, transverse offsets or rotations applied.

Representing the wire array metamaterial in the Drude model, and defining  $\epsilon(\nu) = \epsilon'(\nu)- i\epsilon''(\nu)$ , we have for the real and imaginary components

\begin{subequations}
\label{eq:whole}
\begin{equation}
\epsilon' = 1- \frac{\nu_p^{2}}{\nu^{2}+\Gamma^{2}},
\end{equation}
and
\begin{equation}
\epsilon'' = \left(\frac{\nu_{p}}{\nu}\right)^{2}\frac{\frac{\Gamma}{\nu}}{1+\left(\frac{\Gamma}{\nu}\right)^{2}},
\end{equation}
\end{subequations}

where $\nu$ is the frequency, $\nu_{p}$ is the plasma frequency, and $\mathit{\Gamma}$ the loss term. For normal, or perpendicular incidence, the transmission through a complex dielectric slab is succinctly expressed as  

\begin{equation}
S_{21} = \frac{(1-\rho_{0}^{2})e^{-(\alpha +j\beta)d}}{1-\rho_{0}^{2}e^{-(2\alpha+j\beta)d}},
\end{equation}

where $\rho_{0} =(1-\sqrt{\epsilon})/(1+\sqrt{\epsilon})$, \emph{d} is the width of the array, and $\alpha$ and $\beta$ are the attenuation coefficient and wave numbers of the propagating wave in the medium, respectively [35]. The $\alpha$ and $\beta$ are given by

\begin{subequations}
\label{eq:whole}
\begin{equation}
\alpha(\nu) = \frac{2\pi\nu}{c}\sqrt{\frac{\epsilon'(\nu)}{2}\{\sqrt{1+\rm{tan}^{2}[\delta(\nu)]} -1\}},
\end{equation}
\begin{equation}
\beta(\nu) = \frac{2\pi\nu}{c}\sqrt{\frac{\epsilon'(\nu)}{2}\{\sqrt{1+\rm{tan}^{2}[\delta(\nu)]} +1\}},
\end{equation}
\end{subequations}

and where $\rm{tan} \delta(\nu)=\epsilon'' (\nu)/\epsilon' (\nu)$  is the frequency-dependent loss tangent of the material.  Thus $S_{21}$ data now can be fit with Eq. (4), and metamaterial properties extracted.

The metamaterial properties are encoded in the transition region of the $S_{21}$ around the plasma resonance. Figure 1 shows theoretical transmission versus frequency as a function of the metamaterial characteristics. In each plot, two of three parameters of the array are fixed, with the third varying over a range encompassing representative values for our measurements.

\begin{figure}[ht!]
\centering
\includegraphics[width=.6\textwidth, trim = 6.8cm 3cm 6cm 3cm, clip, scale=1]{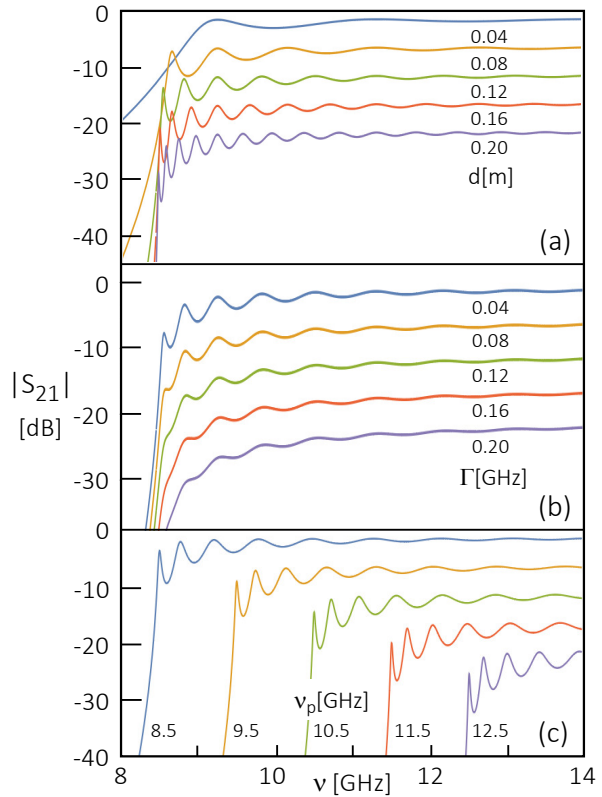}
\caption{\label{fig:epsart} Plots of $|S_{21}|$ where in each case one of the three parameters (plasma frequency, loss rate, width) of the wire array metamaterial is varied in discrete steps, leaving the other two fixed.  (a) Varying the width of the slab \emph{d} [m], with $\nu_{p}$ = 8.5 GHz, $\mathit{\Gamma}$ = 0.01 GHz.  (b) Varying the loss rate $\mathit{\Gamma}$[GHz], with $\nu_{p}$ = 8.5 GHz, \emph{d} = 0.12 m.  (c) Varying the plasma frequency $\nu_{p}$ , with \emph{d} = 0.12 m, $\mathit{\Gamma}$ = 0.01 GHz.  For clarity, the lower curves in (a) are offset in steps of -7.5 dB, for (b) and (c) the offset of the curves are in steps of -5 dB.}
\end{figure}

\subsection{Experimental Setup and Measurements}

The wire planes, 40 in total, consisted of square metal frames of 203 (254) mm inner (outer) edge length, 1.5 mm thickness, and strung with 50 {\textmu}m diameter gold-on-tungsten wire [39] with a wire spacing of \emph{a} = 5.88 mm.  The $S_{21}$ measurements were carried out with standard microwave test equipment, a vector network analyzer (VNA) [41], and matched waveguide horn antennas [42].  

\begin{figure}[ht!]
\centering
\includegraphics[width=80mm, trim = 5cm 3cm 5cm 3cm, clip, scale=0.3]{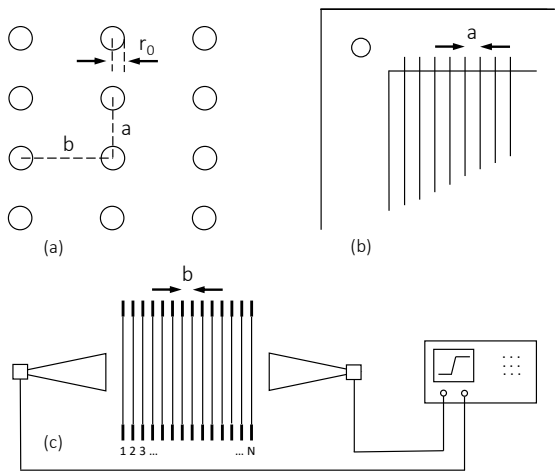}
\caption{\label{fig:epsart} (a)  A rectangular wire array, with $r_{0}$, \emph{a} and \emph{b} designating the radius and the spacing of the wires. (b)  Detail of the construction of a single wire plane.  The wires are supported by the metallic wire frame but not in electrical contact with it.  (c)  The geometry for the measurements of $S_{21}$ as a function of number of wire planes, \emph{N}.}
\end{figure}

The evolution of the plasmonic behavior of the metamaterial as a function of the dimension of the array was first investigated. Individual wire planes were sequentially loaded into a frame, thus constituting a rectangular array with (\emph{a},\emph{b}) = (5.88, 8.00) mm, and the $S_{21}$ measured as a function of the number of wire planes, \emph{N} = 1\,-\,40. This setup is illustrated in Figure 2. The measurements are shown in Figure 3. Even prior to detailed fitting of individual spectra to the model, the data qualitatively exhibit the expected plasmonic behavior.

Repeated measurements were taken under varying conditions to generate a complete data set for fitting $S_{21}$ to the metamaterial properties. These included measurements of different distances between the waveguide horn antennas, and of both uncollimated and collimated geometries. To examine the onset and approach to asymptotic plasmonic response where the fitted parameters changed most rapidly, the low\,-\,\emph{N} spectra (\emph{N} = 1\,-\,20) were repeated separately with a shorter baseline between the horns, and were in good agreement with the \emph{N} = 1\,-\,40 data. The parameters as a function of number of planes \emph{N} = 1\,-\,40 are displayed in Figure 4. For these measurements, the waveguide antennas were spaced 40 cm apart for all \emph{N}. The uncollimated data represents the intrinsic horizontal and vertical beam width of the microwave horns of  $\pm8^{\circ}$ at 3 dB [42]; the collimated data were taken with the same setup, but with the addition of a microwave absorbing collimator in front of the receiver [43] with a rectangular aperture 7.5 cm (horizontal) $\times$ 5.0 cm (vertical). In each case, both collimated and uncollimated data were collected to determine if any scattering from the metal frames was contributing to the spectra; no such evidence was seen. The comparison also confirmed that the data accurately represented strict normal incidence, i.e. the beam perpendicular to the surface of the array as depicted in Fig. 2(c), that is required by Eq. 4, obviating the need to average over the beam spread.

\begin{figure}[ht!]
\centering
\includegraphics[width=.5\textwidth, trim = 8cm 6cm 8cm 5cm, clip, scale=1.0]{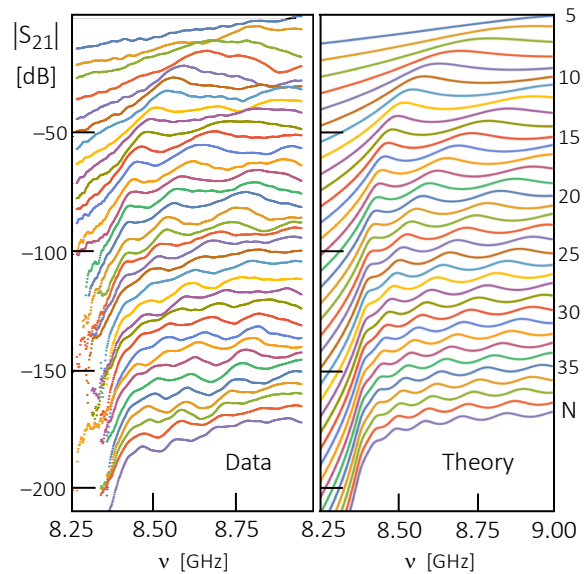}
\caption{\label{fig:epsart} Ensemble of measured and calculated $S_{21}$ for all \emph{N}. For clarity, the spectra are sequentially displaced by –5 dB to avoid overlap, and the data and theory curves are not superposed. To highlight the evolution of the $S_{21}$, the theory spectra are all calculated for $\nu_{p}$ = 8.35 GHz,  $\mathit{\Gamma}$ = 0.04 GHz, and the array width \emph{d} = $Nb$.}
\end{figure}

The data are both presented and fit in terms of the scalar logarithmic gain, i.e. $ g = 20 log_{10} ( | S_{21} | ) $ [dB]. This is particularly important to the fitting, as $S_{21}$ varies steeply in the transition region, which is most sensitive to the metamaterial parameters to be extracted. Furthermore, the fits were most stable and accurate when the fitting region was restricted to a GHz interval around the transition region, beginning just above the noise floor about –40 dB. Including data much above the plateau, $| S_{21} | \sim 1$ did not improve the fit. The resulting fits, now overlaid with the spectra as a function of the number of the planes of the array, \emph{N} are shown in Figure 4.
\begin{figure}[ht!]
\centering
\includegraphics[trim = 8.2cm 5.7cm 7.2cm 5cm, clip, scale=1.15]{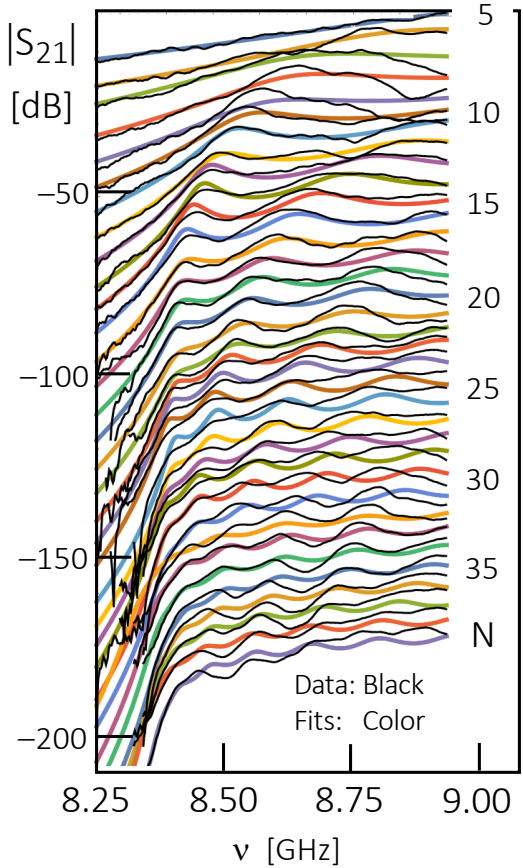}
\caption{\label{fig:epsart}Least-squares fits to spectra spanning the range of the number of wire planes, \emph{N}. For clarity, the spectra are sequentially displaced by –5 dB to avoid overlap.  In all cases, the scale is in absolute units for the topmost curve.}
\end{figure}
We describe here how the data was treated and fit in more detail. Both before and after every series of measurements with the wire planes in place, measurements were made with the planes removed; these two planes-out measurements were averaged, then subtracted from the planes-in measurements, to yield the actual  $S_{21}$ to be fit. While the fact that the data with and without the collimator in place gave consistent results, giving us confidence that the microwave beam was not scattering from the inner edges of the metallic frames, to be absolutely certain that the metallic frames themselves did not contribute in any way to the spectra, one additional check was made. Thus for one series of measurements for \emph{N} = 1-40, we began with 40 dummy frames in place, i.e. with no wires, and then each dummy frame was replaced sequentially with a wire plane. Thus for all spectra \emph{N} = 1-40, there were 40 metallic frames in place, \emph{N} wire loaded and (40 – \emph{N}) empty; these data were indistinguishable from the spectra where no dummy frames were present, only the frames associated with the wire planes themselves.
\section{Results}

The granularity of these data reveals some noteworthy behavior and systematics of wire array metamaterials.  First, the effective width of the full array is equal to its physical width at the few percent level (Figure 5a). It is unambiguously clear, however that the inferred effective width   $d^{\rm{eff}}(N) = 0.00249 + 0.00804 N [\rm{m}]$ is exactly one plane spacing greater than the physical width, $d^{\rm{phy}}(N) = 0.008$(\emph{N}\,-\,1) for all \emph{N}; equivalently it may be stated that the metamaterial effectively extends $\sim$\emph{b}/2 beyond the front and back boundaries of the array.

Second, there is  is already a well-defined plasma frequency even for an array of 5 periods depth, beyond which the plasma frequency increases only marginally; $\nu_{p}$ is within 1\% of its asymptotic value, 8.37 GHz, for 10 periods, and within 0.25\% by 20 periods (Figure 5b).

Third and most importantly, the semianalytic theory of Belov et al. [30] predicts ${\nu_p}^{th}$ = 8.377 GHz for this case, in essentially perfect agreement with the asymptotic value (within 0.1\%).  We have furthermore tested the semianalytic theory as a function of the plane spacing for eight values of $b$, from 3.1 to 7.6 mm, with \emph{N} = 20, for which the agreement on average is within 0.75\% (Figure 6). This is the first precision test of the semianalytic theory.
\begin{figure}[ht!]
\centering
\includegraphics[width=.6\textwidth, trim = 8.45cm 3.5cm 6.25cm 3cm, clip, scale=1.1]{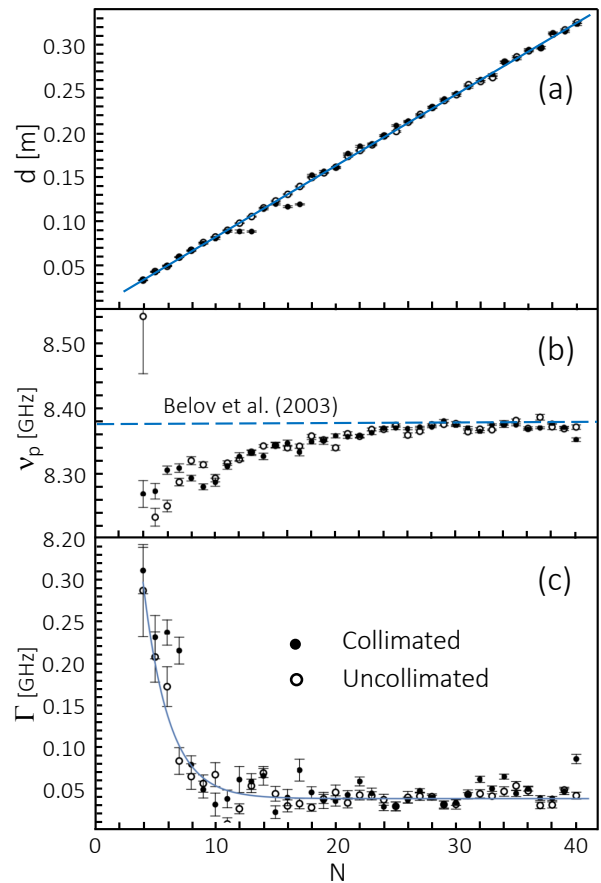}
\caption{\label{fig:epsart}Metamaterial parameters derived from the measured $S_{21}$ as a function of the number of planes, \emph{N}. (a) Effective width of the wire array metamaterial, \emph{d}. (b) Plasma frequency $\nu_{p}$. The blue dashed line indicates the prediction from the semianalytic theory of Ref. [26]. (c) Loss term, $\mathit{\Gamma}$.}
\end{figure}
\begin{figure}[ht!]
\centering
\includegraphics[width=.46\textwidth, trim = 2.8cm 6cm 1.8cm 6cm, clip, scale=.5]{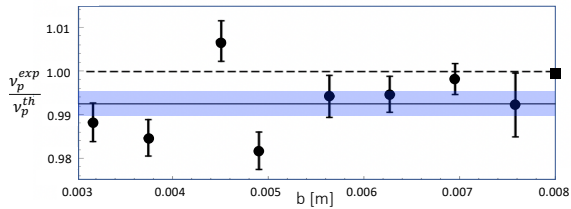}
\caption{\label{fig:epsart}The ratios of the experimental to theoretical plasma frequencies. Closed circles represent values of the plane spacing $b$ = 3.1 – 7.6 mm, with $N$ = 20 planes; the single closed square is for $b$ = 8 mm, with $N$ = 40 planes, where the plasma frequency has reached its asymptotic limit. The solid line represents the average of the $N$ = 20 data points, and the blue band its $\pm 1 \sigma$ error. The theoretical values are from the semianalytic treatment of Belov et al. [30].}
\end{figure}
Finally, an important aim of this study in determining the practical utility of a wire array metamaterial in an axion haloscope was to estimate its quality factor under actual operating conditions, which is directly connected with loss $\mathit{\Gamma}$ extracted from the $S_{21}$. The measured loss exhibits a two-component behavior (Figure 5c), suggestive of a transition from surface-dominated loss (S) to bulk-dominated loss (B). The first reflects the approach to the bulk response,  $\mathit{\Gamma}_{\rm{S}}\cdot \rm{exp}(–\gamma\cdot \emph{N})$, with  $\mathit{\Gamma}_{\rm{S}}$ = 1.70 [GHz], and $\gamma$ = 0.47; the second a constant term  $\mathit{\Gamma}_{\rm{B}}$ = 0.038 [GHz], which dominates for $N > 10$ in the structure.

Some discussion is in order how such wire-array metamaterials would be implemented as a resonator for an axion haloscope. The resonator for an actual haloscope would differ from the open array here in two important ways. The first is in the selection of wire radius and spacing. This study was performed with very thin wires simply for the ease of constructing wire frames and modifying the unit cell over a wide range of configurations, but the wire radius and spacings here would be far from optimal in terms of the resonator's quality factor \emph{Q}. At room temperature and with the current wire parameters, this would imply a quality factor in the bulk limit $Q = \nu_{p} / \mathit{\Gamma} \sim  220$, to be compared with a theoretical value of 260 [44]. However, a resonator of a given frequency can be designed with any choice of correlated wire radius and spacing. These authors find that in the 10 GHz frequency range, the optimum \emph{Q} is achieved for a wire radius of $\sim$ 3 mm, and spacing $\sim$ 10 mm, where the improvement in \emph{Q} would be a factor of $\times$17, bringing the theoretical maximum value to 4,400. Furthermore, as the haloscope would operate at cryogenic temperatures, the \emph{Q} should also improve by the ratio of the classical skin depth at 300 K to the anomalous skin depth below 4 \rm{K}, a factor of $\times$5; values of $\times$3-4 are typical in microwave cavity dark matter experiments. Taken together, one can plausibly project unloaded quality factors $Q > 10^{4}$.

The second important difference is that while the $S_{12}$ measurements to determine metamaterial properties are performed in an open geometry, an actual resonator would consist of the metamaterial enclosed within a microwave cavity, which would support a spectrum of modes appearing as sharp resonances. As shown in Ref. [44], the $TM_{010}$ eigenmode of interest lies slightly higher in frequency than the plasma frequency itself; for simple cavity geometries, the relationship between the eigenfrequency and the plasma frequency can be expressed analytically, but for more complex geometries, by simulation.  

There are two additional requirements for a metamaterial-based haloscope. For an actual axion search, a practical tuning scheme is necessary, ideally one which by modifying the unit cell yields a $\sim$ 30\% dynamic range in frequency the overall volume of the array, but without changing the overall volume of the resonator.  Also needed will be a way of coupling the power out of the resonator, ideally with a single antenna, which can be easily adjusted from weakly to critically coupled.  Designs for such a haloscope resonator are in progress and will be described in a future publication.

\section{Conclusion}
The measurement of the width dependence of wire array metamaterials reveals some novel and surprising behavior, which invites a more detailed theoretical understanding, particularly the early onset of well-defined plasmonic response, and the very different functional approaches of the plasma frequency and the loss term to their respective asymptotic values.

In terms of the potential applicability of wire array metamaterials as an axion haloscope, as proposed in Lawson et al. [28], the semianalytic theory is an excellent tool to guide design, and extrapolated to structures with larger wire radii and spacing, suggest that usable quality factors \emph{Q} are possible.  This of course is just one of several criteria that need to be met, another important question being whether practical dynamic ranges in frequency can be achieved by simple modifications to the unit cell that can be readily engineered; detailed experimental studies of the metamaterial parameters for arrays as a function of unit cell will be reported in the near future.

\textbf{Acknowledgements}.\,-\,This work was supported by the National Science Foundation under Grant No. 1914199. We acknowledge clarifying discussions with M. Lawson, A. Millar, F. Wilczek, P. Belov, R. Balafendiev, and the entire ALPHA collaboration.

\begin{figure}[ht!]
\medskip
\medskip
\textbf{Table of Contents}\\
\medskip
  \includegraphics[width=55mm,height=50mm]{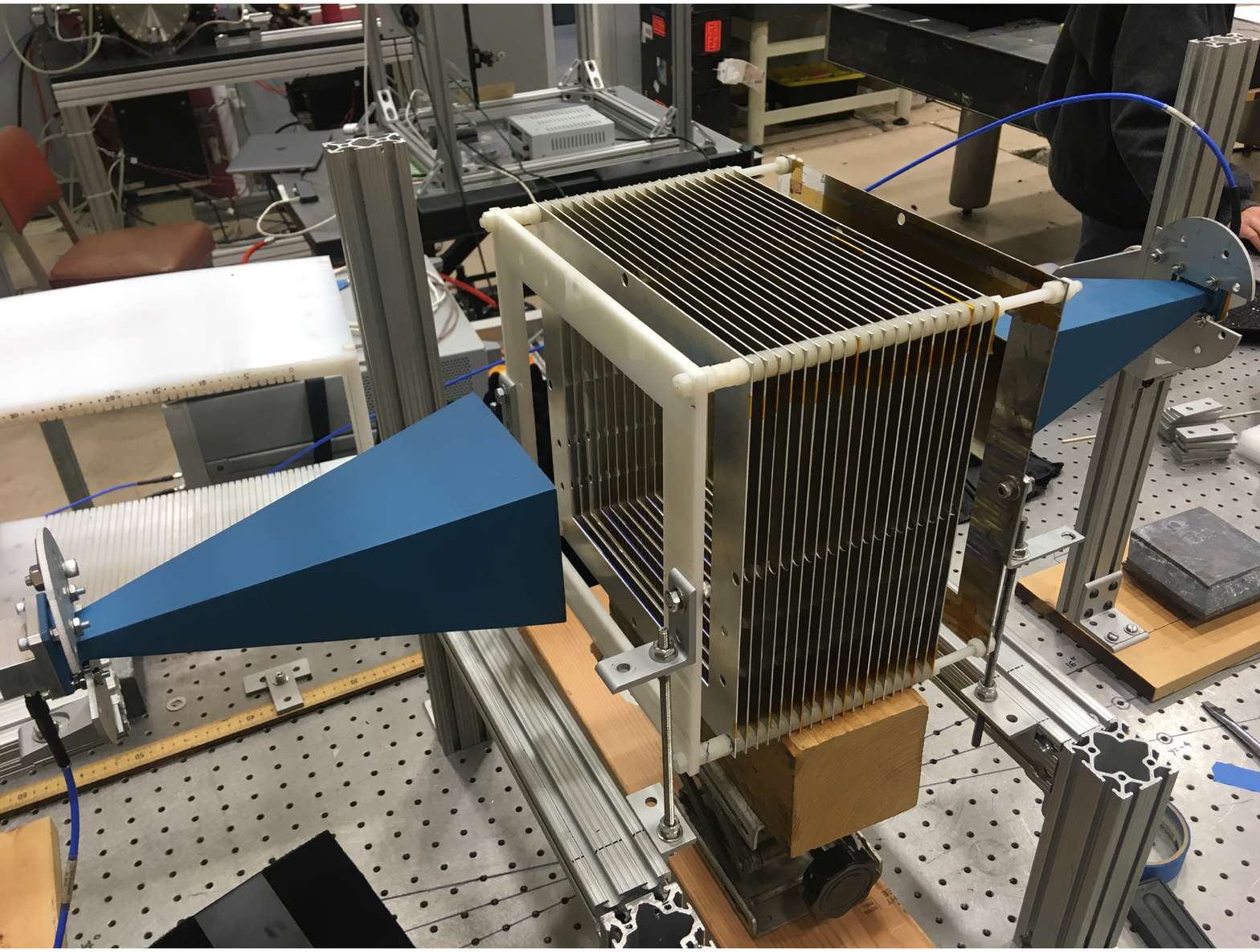}
  \medskip
  \caption*{Wire array metamaterials have been proposed as ideal resonators for dark matter searches at the very high masses predicted for the post-inflation axion, offering the prospect of making resonators simultaneously of very high frequency and very large volume. A detailed exploration of the properties of such wire array metamaterials provides strong support for this concept.}
\end{figure}
\end{document}